\documentclass[journal,onecolumn,11pt,twoside]{IEEEtran}
\usepackage[top=1in, bottom=1in, left=1.25in, right=1.25in]{geometry}
\usepackage{epsfig,graphics,subfigure}
\usepackage{amsmath,amssymb,amsthm,multirow,balance,dsfont,hyperref,cases}
\usepackage{cite,cleveref}
\usepackage[table]{xcolor}
\usepackage{multirow,array}
\usepackage[ruled,vlined]{algorithm2e}
\usepackage{colortbl}
\usepackage{rotating} 

\usepackage{empheq}
\usepackage[most]{tcolorbox}

\newtcbox{\mymath}[1][]{%
    nobeforeafter, math upper, tcbox raise base,
    enhanced, colframe=blue!30!black,
    colback=blue!30, boxrule=0.5pt,size=small,
    #1}

\usepackage{scalerel,stackengine}
\stackMath
\newcommand\widecheck[1]{%
\savestack{\tmpbox}{\stretchto{%
  \scaleto{%
    \scalerel*[\widthof{\ensuremath{#1}}]{\kern-.3pt\bigwedge\kern-.3pt}%
    {\rule[-\textheight/3]{1ex}{\textheight}}
  }{\textheight}%
}{0.5ex}}%
\stackon[1pt]{#1}{\scalebox{-0.8}{\tmpbox}}%
}

\newcommand{\vertiii}[1]{{\left\vert\kern-0.25ex\left\vert\kern-0.25ex\left\vert #1 
    \right\vert\kern-0.25ex\right\vert\kern-0.25ex\right\vert}}

\newtheorem{example}{Example}

\newtheorem{remark}{Performance result}

\usepackage{color}
\def\cblue{\textcolor{black}}

\definecolor{beaublue}{rgb}{0.84, 0.89, 1}

\newcommand{\boldh}{\boldsymbol{h}}

\newcommand{\bd}{\boldsymbol{d}}
\newcommand{\bs}{\boldsymbol{s}}
\newcommand{\bu}{\boldsymbol{u}}
\newcommand{\bv}{\boldsymbol{v}}
\newcommand{\bw}{\boldsymbol{w}}

\newcommand{\bx}{\boldsymbol{x}}

\newcommand{\bgamma}{\boldsymbol{\gamma}}
\newcommand{\bpsi}{\boldsymbol{\psi}}
\newcommand{\bphi}{\boldsymbol{\phi}}

\newcommand{\cA}{\mathcal{A}}

\newcommand{\cC}{\mathcal{C}}

\newcommand{\cR}{\mathcal{R}}

\newcommand{\cL}{\mathcal{L}}

\newcommand{\cN}{\mathcal{N}}
\newcommand{\cP}{\mathcal{P}}

\newcommand{\cU}{\mathcal{U}}
\newcommand{\cu}{{\scriptscriptstyle\mathcal{U}}}

\newcommand{\cw}{{\scriptstyle\mathcal{W}}}

\newcommand{\ccw}{{\scriptscriptstyle\mathcal{W}}}

\newcommand{\bcw}{\boldsymbol{\cw}}

\newcommand{\cwb}{\overline{\cw}}
\newcommand{\wb}{\overline{w}}

\newcommand{\expec}{\mathbb{E}}

\newcommand{\col}{\text{col}}

\newcommand{\prox}{\text{prox}}

\newcommand{\diag}{\text{diag}}
\newcommand{\sign}{\text{sign}}
\DeclareMathOperator*{\argmin}{argmin}

\DeclareMathOperator*{\st}{subject~to}

\renewcommand{\IEEEbibitemsep}{0pt plus 0.5pt}
\makeatletter
\IEEEtriggercmd{\reset@font\normalfont\fontsize{8.8pt}{9.3pt}\selectfont}
\makeatother
\IEEEtriggeratref{1}

\newcolumntype{C}[1]{>{\centering\arraybackslash}m{#1}}
\begin{document}
\title{Multitask learning over graphs: \\An Approach for Distributed, Streaming Machine Learning\vspace{-3mm}}
\author{Roula Nassif, Stefan Vlaski, C\'edric Richard, Jie Chen,  and Ali H. Sayed\\
\thanks{ 
\cblue{This work was submitted while R. Nassif was a post-doc at EPFL. She is now with the American University of Beirut, Lebanon (e-mail: roula.nassif@aub.edu.lb).} S. Vlaski and A. H. Sayed are with Institute of Electrical Engineering, EPFL, Switzerland (e-mail: $\{$stefan.vlaski, ali.sayed$\}$@epfl.ch). C. Richard is with Universit\'e de Nice Sophia-Antipolis, France (cedric.richard@unice.fr). J. Chen is with Northwestern Polytechnical~University,~China~(dr.jie.chen@ieee.org).}
}

\maketitle
\vspace{-1.5cm}
\begin{abstract}
\vspace{-4mm}
The problem of learning simultaneously several related tasks has received considerable attention in several domains, especially in machine learning with the so-called \emph{multitask learning} problem or \emph{learning to learn} problem~\cite{caruana1997multitask,thrun1998learning}. 
Multitask learning is an approach to inductive {transfer learning} (using what is learned for one problem to {assist in} another problem) and helps improve generalization performance relative to learning each task separately by using the domain information contained in the training signals of \emph{related} tasks as an inductive bias. Several strategies {have been} derived within this community under the assumption that all data are available beforehand at a fusion center.  {However, recent} years have witnessed an increasing ability to collect data in a distributed and streaming {manner}. 
This requires the design of new strategies for learning jointly multiple tasks  from streaming data  over distributed (or networked) systems. This article provides an overview of multitask strategies for learning and adaptation over networks.  The working hypothesis for these strategies is that agents are allowed to cooperate with each other in order to learn distinct, though related tasks. The article shows how cooperation steers the network limiting point and how different cooperation rules allow to promote different task relatedness models. It also explains how and when cooperation over multitask networks outperforms non-cooperative strategies. 
\end{abstract}
\vspace{-0.5cm}

\section{Multitask network models}
\label{sec: Multitask network models}
\begin{figure*}[t]
\centering
\includegraphics[scale=0.33]{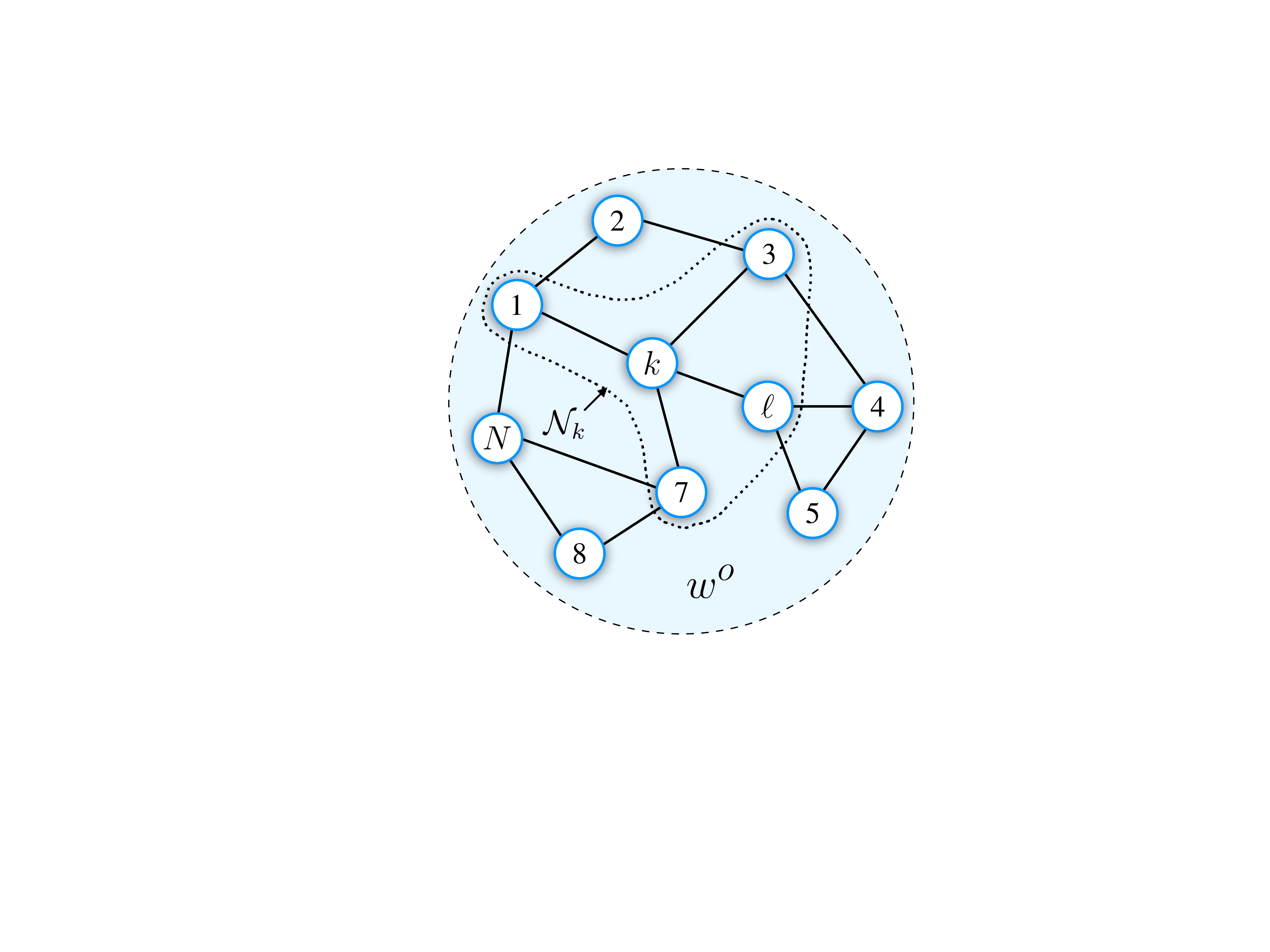}
\includegraphics[scale=0.33]{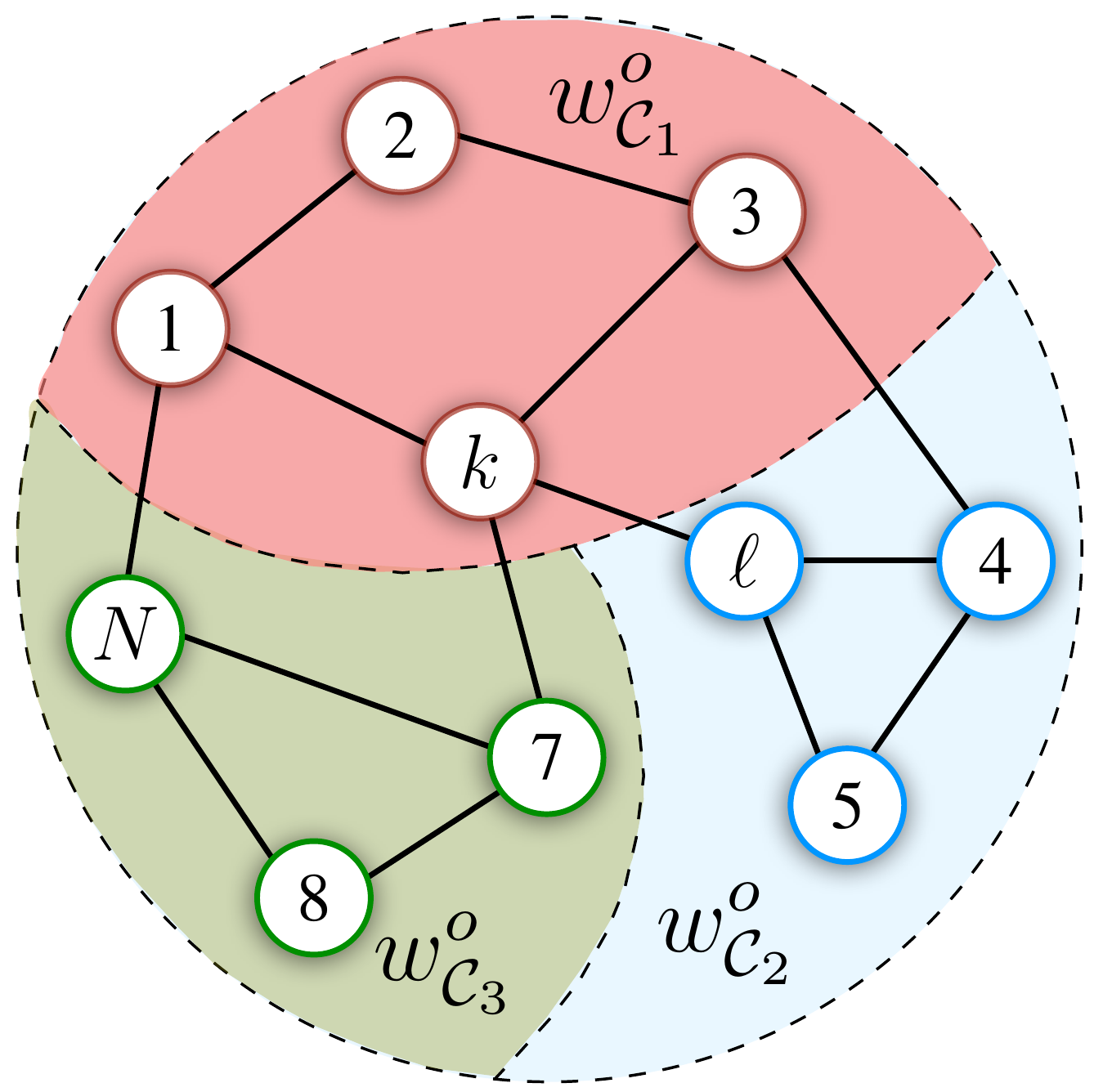}
\includegraphics[scale=0.33]{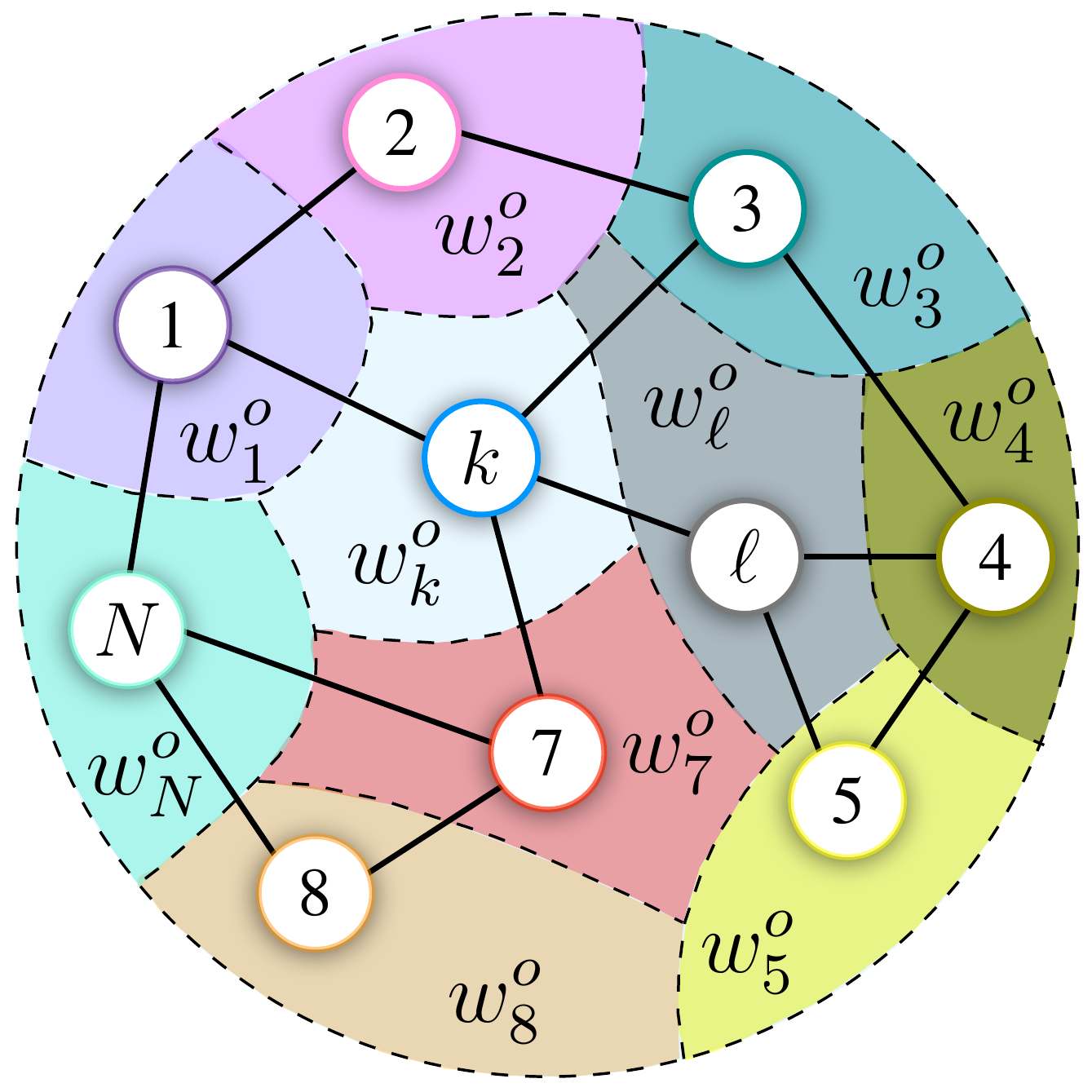}
\vspace{-2mm}
\caption{Network models. \textit{(Left)} Single-task network. \textit{(Middle)} Clustered multitask network. \textit{(Right)} Multitask network.}
\label{fig: multitask network models}
\vspace{-3mm}
\end{figure*}
Consider a networked system consisting of a collection of $N$ autonomous agents (sensors, classifiers, etc.) distributed over some geographic area and connected through a  topology. The neighborhood of agent $k$ is denoted by $\cN_k$; it consists of all agents that are connected to $k$ by an edge--see Fig.~\ref{fig: multitask network models} (left). A real-valued strongly convex and differentiable cost  $J_k(w_k)$ is associated with each agent $k$. The objective (or the task) at agent $k$ is to estimate the parameter vector, $w^o_k$, of size $M_k\times 1$, that minimizes $J_k(w_k)$, namely,
\begin{equation}
\label{eq: definition of the task}
w^o_k\triangleq\arg\min_{w_k}J_k(w_k).
\end{equation} 
Depending on how the minimizers across the agents relate to each other, we distinguish between three categories of networks:
\begin{enumerate}
\item \textit{Single-task network:} All costs $J_k(w_k)$ are minimized at the same location $w^o$, namely, $w^o_k=w^o$ for all $k$ -- see Fig.~\ref{fig: multitask network models} (left). 
\item \textit{Clustered multitask network:} The $N$ agents are grouped into $Q$ clusters $\cC_q$ ($q=1,\ldots,Q$) and, within each cluster $\cC_q$, all the costs are minimized at the same location $w^o_{\cC_q}$, namely, $w^o_k=w^o_{\cC_q}$ for all $k\in\cC_q$ -- see Fig.~\ref{fig: multitask network models} (middle). 
Similarities or relationships may exist among the distinct minimizers $\{w^o_{\cC_q}\}$.
\item \textit{Multitask network:}  The individual costs are minimized at distinct, though related, locations $\{w^o_k\}$ -- see Fig.~\ref{fig: multitask network models} (right). 
\end{enumerate}
Each agent $k$ can solve~\eqref{eq: definition of the task} on its own. However, since the objectives across the network relate to each other, it is expected that by properly promoting these relationships, one may improve the network performance. In other words, it is expected that {through cooperation among the agents}, one may improve the network performance. One important question is how to design cooperative strategies that can lead to better performance than non-cooperative ones where each agent {attempts} to determine $w^o_k$ on its own. This overview paper explains how multitask learning over graphs addresses this question.

\cblue{Prior to multitask learning over graphs, there have been many works in the machine learning literature where learning multiple related tasks simultaneously has been considered~\cite{caruana1997multitask,thrun1998learning,zhang2010convex,chen2010graphstructured,evgeniou2004regularized,jacob2008clustered,kato2007multitask}. Multitask learning was shown, both empirically and theoretically, to improve performance relative to the traditional approach of learning each task separately
. Depending on the machine learning application, several task relatedness models have been considered. For example, in~\cite{caruana1997multitask,evgeniou2004regularized}, the functions to be learned are assumed to share a common underlying representation. In~\cite{jacob2008clustered}, it is assumed that the tasks are close to each other in some Hilbert space. 
Probabilistic based approaches, where a probability model capturing the relations between tasks is estimated simultaneously with functions corresponding to each task,  have also been considered~\cite{zhang2010convex}. Also, graph-based approaches, where the relations between tasks are captured by an underlying graph, were also considered in the literature~\cite{chen2010graphstructured,kato2007multitask}. All these works, however, assume that all data  are available beforehand at a fusion center and propose batch-mode methods to solve multitask problems. Other existing works, such as~\cite{wang2016distributed}, consider distributed data setting. However, most of these works still require an architecture consisting of workers along with a master, where agents perform local computations followed by sending intermediate results to the master for further processing. Such solution methods are not fully distributed, which limits their range~of~practical~applications.}

\cblue{This paper, however, focuses on fully distributed solutions that avoid the need for central data aggregation or processing and instead rely on local computations and communication exchanges among neighborhoods. Besides providing distributed implementations, the solutions considered in this paper are able to learn continuously from streaming data. We start our exposition by describing a class of non-cooperative solutions that are able to respond in real time to streaming data. Then, we explain how these solutions can be extended to handle multitask learning over~graphs.}

\vspace{-2.5mm}
\section{Noncooperative {learning under} streaming data}
\vspace{-1mm}
\label{sec: Streaming data setting}
Throughout this article, there is an explicit assumption that agents operate \cblue{in the} streaming data setting. That is, it is assumed that each {agent $k$} receives at each time instant $i$ \cblue{one instantaneous realization $\bx_{k,i}$ of a random data  $\bx_k$. }
The goal of agent $k$ is to estimate the vector $w^o_k$ that minimizes {its} risk \cblue{function $J_k(w_k)\triangleq \expec_{\bx_k} Q_k(w_k;\bx_{k})$}, defined in terms of some loss function $Q_k(\cdot)$. The expectation is computed over the distribution of the \cblue{data $\bx_k$}. Agent $k$ is particularly interested in solving the problem in the \emph{stochastic} setting when the distribution of the data is generally unknown. This means that the risks $J_k(\cdot)$ and their gradients $\nabla_{w_k}J_k(\cdot)$ are unknown. As such{,} approximate gradient vectors $\widehat{\nabla_{w_k}J_k}(\cdot)$ {will need} to be employed. Doing so leads to the following stochastic gradient algorithm for solving~\eqref{eq: definition of the task}:
\vspace{-1mm}
\begin{equation}
\label{eq: non-cooperative strategy}
\bw_{k,i}=\bw_{k,i-1}-\mu\widehat{\nabla_{w_k}J_k}(\bw_{k,i-1}),
\vspace{-2mm}
\end{equation}
where $\bw_{k,i}$ is the estimate of $w^o_k$ at iteration $i$ and $\mu>0$ is a small step-size parameter. \cblue{Resorting to the instantaneous realization $\bx_{k,i}$ of the random data $\bx_k$,  a} common construction in the stochastic approximation theory is to employ the {following gradient} approximation at iteration $i$:
\begin{equation}
\label{eq: approximation gradient}
\widehat{\nabla_{w_k}J_k}(w_k)=\nabla_{w_k}Q_k(w_k;\bx_{k,i}).
\end{equation}
We therefore focus in this paper on stochastic gradient algorithms{,} which are powerful iterative procedures for solving~\eqref{eq: definition of the task} in the streaming data. They enable continuous learning and adaptation in response to drifts in the location of the minimizers due to changes in the costs. We illustrate construction~\eqref{eq: non-cooperative strategy}--\eqref{eq: approximation gradient} by considering scenarios from machine learning and adaptive filter theory.
\begin{example}{\emph{(Logistic regression network).}} 
\label{exple: logistic regression}
\emph{Let $\bgamma_k(i)=\pm 1$ be a streaming sequence of (class) binary random variables and let $\boldh_{k,i}$ be the corresponding streaming sequence of $M_k\times 1$ real random (feature) vectors with $R_{h,k}=\expec\boldh_{k,i}\boldh_{k,i}^\top>0$. The processes $\{\bgamma_k(i),\boldh_{k,i}\}$ are assumed to be wide-sense stationary. In these problems, agent $k$ seeks to estimate the vector $w^o_k$ that minimizes the regularized logistic risk function~\cite{sayed2014adaptation}:
\begin{equation}
\label{eq: logistic regression cost}
J_k(w_k)=\expec\ln\left(1+e^{-\bgamma_k(i)\boldh_{k,i}^\top w_k}\right)+\frac{\rho}{{2}}\|w_k\|^2,
\end{equation}
where $\rho>0$ is a regularization parameter. Once $w^o_k$ is found, $\widehat{\gamma}_k(i)=\sign(\boldh_{k,i}^\top w^o_k)$ can then be used as a decision rule to classify new features. Using approximation~\eqref{eq: approximation gradient}, we obtain the following stochastic-gradient algorithm for minimizing~\eqref{eq: logistic regression cost}:
\begin{equation}
\bw_{k,i}=(1-\mu\rho)\bw_{k,i-1}+\mu\bgamma_k(i)\boldh_{k,i}\left(\frac{1}{1+e^{\bgamma_k(i)\boldh_{k,i}^\top\bw_{k,i-1}}}\right).
\end{equation}}\qed
\end{example}
\begin{example}{\emph{(Mean-square-error (MSE) network).}} 
\label{exple: mse network}
\emph{In such networks, each agent is subjected to streaming data $\{\bd_k(i),\bu_{k,i}\}$ that are assumed to satisfy a linear regression model:
\begin{equation}
\label{eq: linear data models}
\bd_{k}(i)=\bu_{k,i}^\top w^o_k+\bv_k(i),
\end{equation}
for some unknown $M_k\times 1$ vector $w^o_k$ to be estimated by agent $k$ with $\bv_k(i)$ denoting a zero-mean measurement noise. For these networks, the risk function takes the form of \cblue{{an} MSE} cost~\cite{sayed2013diffusion}:
\begin{equation}
\vspace{-1mm}
\label{eq: mean-square-error}
J_k(w_k)=\frac{1}{2}\expec\left(\bd_{k}(i)-\bu_{k,i}^\top w_k\right)^2,
\end{equation}
which is minimized at $w^o_k$. The processes $\{\bu_{k,i},\bv_k(i)\}$ are zero-mean jointly wide-sense stationary with: i) $\expec\bu_{k,i}\bu_{\ell,i}^\top=R_{u,k}>0$ if $k=\ell$ and zero otherwise; $\expec\bv_{k}(i)\bv_{\ell}(i)=\sigma^2_{v,k}$ if $k=\ell$ and zero otherwise; and iii) $\bu_{k,i}$ and $\bv_{k}({j})$ are independent of each other. Using approximation~\eqref{eq: approximation gradient}, we obtain the following stochastic-gradient algorithm:
\begin{equation}
\vspace{-1mm}
\label{eq: lms algorithm}
\bw_{k,i}=\bw_{k,i-1}+\mu\bu_{k,i}(\bd_{k}(i)-\bu_{k,i}^\top\bw_{k,i-1}),
\end{equation}
which is the well-known least-mean-squares (LMS) algorithm~\cite{widrow1985adaptive}.}\qed
\end{example}

The use of {the approximate} gradient $\widehat{\nabla_{w_k}J_k}(\cdot)$ instead of the true gradient  ${\nabla_{w_k}J_k}(\cdot)$ in~\eqref{eq: non-cooperative strategy} introduces perturbations into the operation of the gradient descent iteration. This perturbation is referred to as the gradient noise defined as 
$\bs_{k,i}(w_k)\triangleq{\nabla_{w_k}J_k}(w_k)-\widehat{\nabla_{w_k}J_k}(w_k).$
The presence of this perturbation prevents the stochastic iterate $\bw_{k,i}$ from converging almost surely to the minimizer $w^o_k$ when constant step-sizes are used. Some deterioration in performance will occur, and the iterate $\bw_{k,i}$ will instead {fluctuate} close to $w^o_k$. It is common \cblue{in adaptive filtering and stochastic gradient optimization literatures} to assess the size of these fluctuations by measuring their steady-state mean-square \cblue{value~\cite{widrow1985adaptive,sayed2013diffusion,sayed2014adaptation}}. We therefore focus in this paper on highlighting the benefit of multitask learning on the network mean-square-deviation (MSD), which is defined as the steady-state average variance value:
\begin{equation}
\vspace{-0.6mm}
\label{eq: network MSD}
\text{MSD}\triangleq\lim_{i\rightarrow \infty}\frac{1}{N} \sum_{k=1}^N\expec\|w^o_k-\bw_{k,i}\|^2.
\end{equation} 
 In the sequel, when discussing theoretical performance results, and to avoid excessive technicalities, it is sufficient to focus on MSE networks described in Example~\ref{exple: mse network} and to assume that $R_{u,k}=R_u$ and $M_k=M$ for all $k$\footnote{Performance results under more general conditions, such as allowing for space dependent covariances and lengths and for general \cblue{second-order differentiable} cost functions that are not necessarily quadratic, can also be found in~\cite[Chap.~3--4]{sayed2014adaptation} for algorithm~\eqref{eq: non-cooperative strategy}, in~\cite{nassif2018diffusion} for the strategy introduced in Sec.~\ref{sec: Multitask estimation under smoothness}, and in~\cite{nassif2019adaptation2} for the strategies in Sec.~\ref{sec: Multitask estimation under subspace constraints}. \cblue{The MSD performance expressions in these works are derived under Lipschitz gradient vectors and Hessian matrices assumptions. It should be noted that the analyses in these works allow also to recover the Excess-Risk metric at agent $k$, which is defined as $\text{ER}_k\triangleq\lim_{i\rightarrow\infty}\expec(J_k(\bw_{k,i})-J_k(w^o_k))$--see, e.g.,~\cite[p. 388--390]{sayed2014adaptation}. Due to space limitations, we shall only focus on presenting MSD performance expressions.}}. In this way, the quality of the measurements, captured by the noise power $\sigma^2_{v,k}$, is allowed to vary across the network with some agents collecting noisier data than other agents. Assuming uniform regressors covariance allows us to quantify the improvement in performance that results from cooperation without biasing the results by the statistical nature of the regression data at the agents. 
\vspace{-0.1cm}
\begin{remark}
\label{remark: performance of non-cooperative strategy}
Consider \cblue{an MSE} network running the non-cooperative algorithm~\eqref{eq: lms algorithm}. Assume further that  $R_{u,k}=R_u$ and $M_k=M$ for all $k$. Under these assumptions, and for sufficiently small step-sizes, the individual steady-state variance $\emph{MSD}_k\triangleq\lim_{i\rightarrow\infty}\expec\|w^o_k-\bw_{k,i}\|^2$ and the network \emph{MSD} defined by~\eqref{eq: network MSD} are given by~\cite{sayed2013diffusion}:
\begin{equation}
\vspace{-2mm}
\label{eq: non-cooperative performance}
\emph{MSD}_k=\frac{\mu M}{2}\cdot\sigma^2_{v,k},\qquad\emph{MSD}^\emph{nc}=\frac{\mu M}{2}\cdot\left(\frac{1}{N}\sum_{k=1}^N\sigma^2_{v,k}\right),
\end{equation}\qed\vspace{-2mm}
\end{remark}
\noindent \cblue{where the superscript ``nc'' is used to indicate that the MSD expression is for the non-cooperative solution.} First, observe that the performance is on the order of $\mu$. The smaller $\mu$ is, the better the performance will be, but the slower the convergence toward $w^o_k$ will be~\cite{sayed2014adaptation,sayed2013diffusion} (the same observation is valid for future expressions~\eqref{eq: steady state variance smoothness} and~\eqref{eq: projection performance} with convergence to $\cw^\star$ in~\eqref{eq: steady state variance smoothness} instead). Second, observe that agents with noisier data will perform worse than agents with cleaner data.  However, since agents are observing data arising from similar or related models $w^o_k$,  it is expected that an appropriate cooperation among agents can help enhance the network performance.
\vspace{-2mm}
\section{Multitask learning framework}
Depending on the application, several task relatedness models {can be considered}. For each  model, an appropriate  convex optimization problem {is} solved in a distributed and adaptive manner. 
This results in different multitask strategies, and therefore different cooperation rules between agents.  Rather than describing each optimization problem in isolation, we begin by introducing a general problem, which will allow us to recover various multitask strategies as special cases.

Let $\cw\triangleq \col\{w_1,\ldots,w_N\}$ denote the collection of parameter vectors from across the network. We consider the following global optimization problem for the multitask formulation:
\vspace{-1.5mm}
\begin{empheq}[box={\mymath[colback=blue!5,drop lifted shadow, sharp corners]}]{equation}
\label{eq: multitask optimization problem}
\begin{split}
\cw^\star=&~\arg\min_{\ccw}~~J^\text{glob}(\cw)=\sum_{k=1}^NJ_k(w_k)+{\frac{\eta}{2}} \cR(\cw)\\
&~\st~\cw\in\Omega
\end{split}
\end{empheq}
where $\cR(\cdot)$ is a convex regularization function promoting the relationships between the tasks, $\Omega$ is a closed convex set defining the feasible region of the parameter vectors, and $\eta>0$ is a parameter controlling the importance of the regularization. The choice of the regularizer $\cR(\cdot)$ and the set $\Omega$ depends on the prior information on how the multitask models relate to each other. {To illustrate how problem formulation~\eqref{eq: multitask optimization problem} can be used, we consider the following two examples that are multitask oriented.}

\begin{figure}
\centering
\includegraphics[scale=0.28]{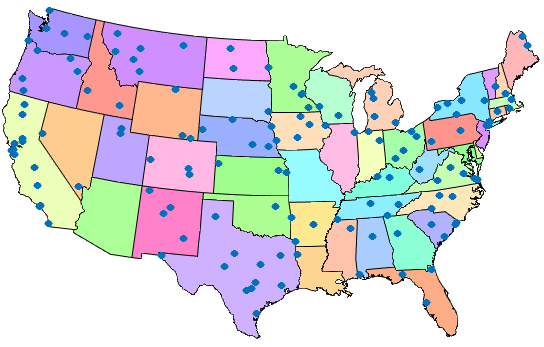}\quad
\includegraphics[scale=0.25]{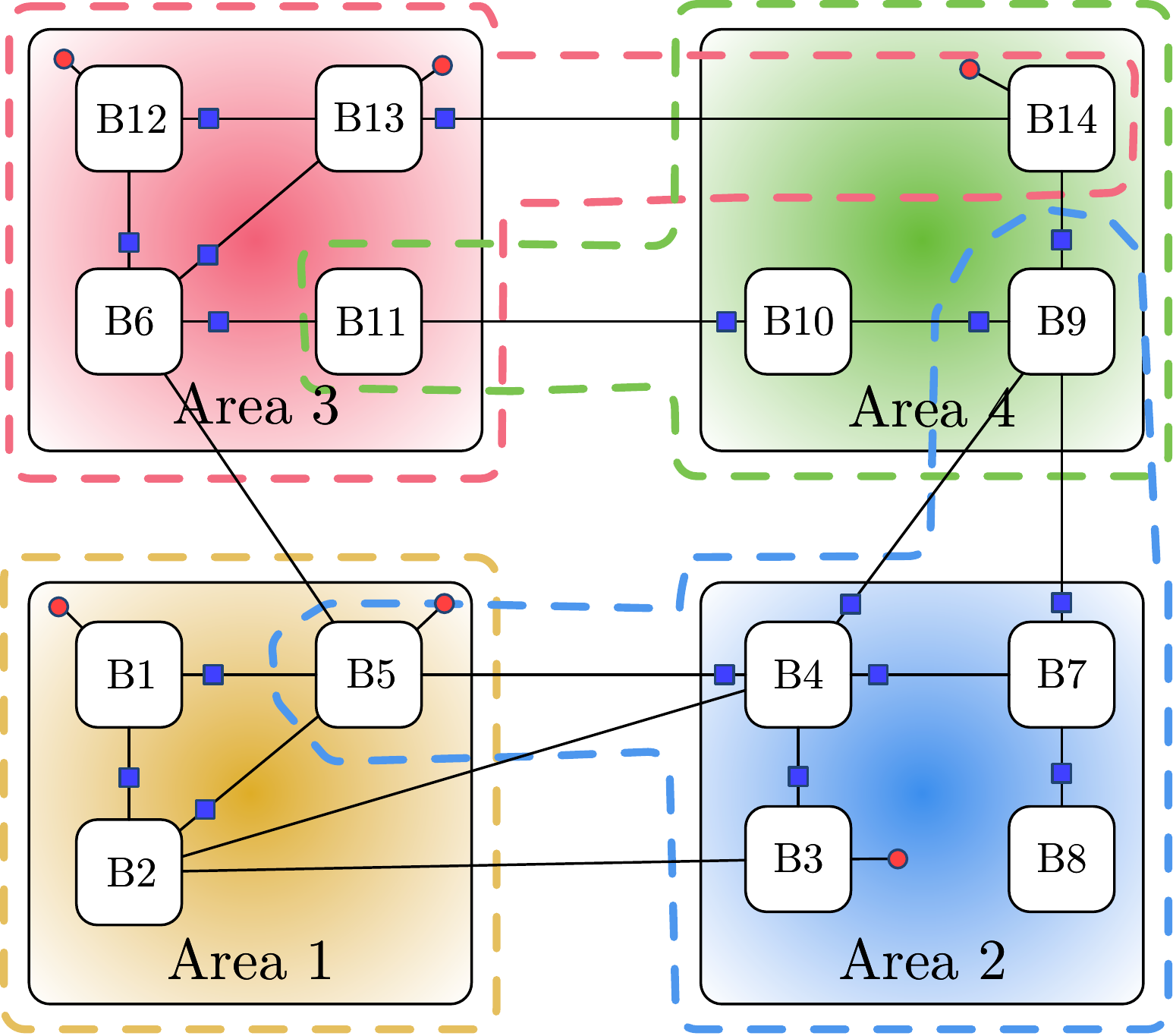}
\vspace{-3mm}
\caption{Examples of multitask applications. \textit{(Left)} Weather forecasting. \textit{(Right)} Distributed power system monitoring.}
\label{fig: multitask network applications}
\vspace{-3mm}
\end{figure}
\vspace{-2mm}
\begin{example}{\emph{(Weather forecasting~\cite{nassif2018diffusion}).}} 
\label{exple: weather forecasting}
\emph{Consider the network in Fig.~\ref{fig: multitask network applications} (left) consisting of $N=139$ weather stations located {across} the United States and collecting daily measurements
. Let $\boldh_{k,i}$ denote the feature vector consisting of collected data (temperature, wind speed, dew point, etc.) at sensor $k$ and day $i$ and let $\bgamma_k(i)$ denote the corresponding binary variable associated with rain occurrence, i.e.,  $\bgamma_{k}(i) = 1$ if rain occurred and  $\bgamma_{k}(i) = -1$ otherwise. The objective is to construct a classifier at each station to predict whether it will rain or not based on the knowledge of $\boldh_{k,i}$. To this end, each station can use an individual logistic regression machine {similar to} the one described in Example~\ref{exple: logistic regression}; {in this case the cost $J_k(w_k)$ in~\eqref{eq: multitask optimization problem} takes the form~\eqref{eq: logistic regression cost}}. However, it is expected that the decision rules $\{w^o_k\}$ at neighboring stations {would be} similar since they are collecting features arising from similar statistical distributions. Moreover, the strength of  similarity is expected to be inversely proportional to the physical distance between the stations. This  gives rise to a weighted graph {(with closest nodes connected by edges)} and one may expect to improve the network performance by promoting the smoothness of $\{w^o_k\}$ with respect to the underlying graph. {The simplest possible term that encourages smoothness is the graph Laplacian regularizer $S(\cw)$ defined further ahead in~\eqref{eq: smoothness measure}}. {By choosing $\cR(\cw)=S(\cw)$ and $\Omega=\mathbb{R}^{MN}$ in~\eqref{eq: multitask optimization problem}, one arrives at a multitask formulation for the weather forecasting application that takes into account the smoothness prior over the graph.} {This formulation and other possible formulations  are solved in Sections~\ref{sec: Regularized multitask estimation} and~\ref{sec: Multitask estimation under subspace constraints}. 
}\qed}
\end{example}

\vspace{-5mm}
\begin{example}{\emph{(Power system state monitoring~\cite{kekatos2013distributed}).}}
\label{exple: Power system state monitoring}
\emph{Consider Fig.~\ref{fig: multitask network applications} (right) illustrating an IEEE 14-bus power monitoring system partitioned into 4 areas,  where each area comprises a subset of buses supervised by its own control center. The local state vectors (bus voltages) to be estimated at neighboring areas may partially overlap as the areas are interconnected. This is because each control center collects measurements related to the voltages across its  local  buses and voltages across  the interconnection between neighboring centers. For example, Area 2 supervises buses 3, 4, 7, and 8. Since it collects current readings on lines (4, 5) and (7, 9), its state vector  extends to buses 5 (supervised by Area 1) and 9 (supervised by Area 4). {In other words, if we let $w^n$ denote the state of bus $n$, then the cost $J_2(\cdot)$ at Area 2 will depend on  the extended parameter vector $w_2=\col\{w^3,w^4,w^5,w^7,w^8,w^9\}$. However, since the parameter vectors at Areas 1 and 4 will be $w_1=\col\{w^1,w^2,w^5\}$ and $w_4=\col\{w^9,w^{10},w^{11},w^{14}\}$, respectively, consensus needs to be reached on the variable $w^5$ between Areas 2 and 1, and on the variable $w^9$ between Areas 2 and 4, 
while minimizing the individual cost $J_2(w_2)$ penalizing deviation from {data models of the form $y_{k}=H_{k}w_k+v_{k}$ where $H_{k}$ is the measurement matrix and $v_{k}$ is a zero-mean  noise. Thus, distributed power state estimation can be formulated as problem~\eqref{eq: multitask optimization problem} with $\cR(\cw)=0$, whereas the constraint set $\Omega$ in this case should be selected to promote consensus over the overlapped variables. In Section~\ref{subsec: Multitask estimation with overlapping parameter vectors}, we explain how such  problems can be solved.}} \qed
} 
\end{example}

{Returning to the formulation~\eqref{eq: multitask optimization problem}, observe that} even though the aggregate cost $\sum_{k=1}^NJ_k(w_k)$ is separable in $w_k$, the cooperation between agents is necessary due to the coupling between the tasks through the regularization and the constraint. Note that, when solving problem~\eqref{eq: multitask optimization problem}, agent $k$ will be responsible {for} estimating $w_k^\star$ (the $k$-th sub-vector of $\cw^\star= \col\{w_1^\star,\ldots,w_N^\star\}$), which {is generally} different {from} $w^o_k$ in~\eqref{eq: definition of the task}, the actual objective at agent $k$.  However, it is expected that accurate prior information  will allow the designer to choose the regularizer $\cR(\cdot)$, the set $\Omega$, and the strength $\eta$ in a way that minimizes the distance between $w^\star_k$ and $w^o_k$.

\begin{figure}
\centering
\includegraphics[scale=0.33]{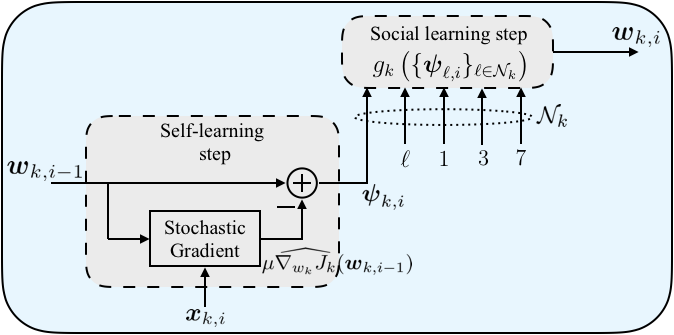}
\vspace{-2mm}
\caption{A common diagram for the multitask strategies described in this work. The structure involves two main steps: i) a self-learning step~\eqref{eq: self-learning step}, and ii) a social learning step~\eqref{eq: social learning step}.}
\vspace{-2mm}
\label{fig: multitask diagram}
\end{figure}
Although some existing works use primal-dual \cblue{methods~\cite{koppel2016proximity}} to solve multitask estimation problems, we limit our exposition to the class of primal techniques (based on propagating and estimating the primal variable) that employ \emph{stochastic-gradient}  iterations. {Extensive studies in the literature have shown that small step-sizes enable these strategies to learn well in \emph{streaming} data settings.} Due to the separability property of $\sum_{k=1}^NJ_k(w_k)$, the multitask  algorithms described in the {sequel} will have a common structure given by:
\vspace{-2mm}
\begin{subequations}
\label{eq: multitask algorithm}
\begin{empheq}[box={\mymath[colback=beaublue,drop lifted shadow, sharp corners]}]{align}
\bpsi_{k,i}&=\bw_{k,i-1}-\mu\widehat{\nabla_{w_k}J_k}(\bw_{k,i-1})\label{eq: self-learning step}\\
\bw_{k,i}&=g_k\left(\{\bpsi_{\ell,i}\}_{\ell\in\cN_k}\right)\label{eq: social learning step}
 \end{empheq}
\end{subequations}
The first step~\eqref{eq: self-learning step} corresponds to the stochastic gradient step on the individual cost $J_k(\cdot)$. We refer to this step as the \emph{self-learning} step--see Fig.~\ref{fig: multitask diagram}. Compared with the non-cooperative strategy~\eqref{eq: non-cooperative strategy}, observe now that the result of the gradient descent step is $\bpsi_{k,i}$, an intermediate estimate of $w^o_k$ at iteration $i$. This step {is} followed by a \emph{social learning} step~\eqref{eq: social learning step}{, which uses some function $g_k(\cdot)$ of the neighborhood iterates.} As we shall see in the {next sections}, the {form} of this function depends on the regularizer $\eta \cR(\cdot)$ and the set $\Omega$ in~\eqref{eq: multitask optimization problem}, {both of which} allow to promote the prior information on how the tasks $w^o_k$ are related.
The result of this second step is $\bw_{k,i}$, the estimate of $w^o_k$, defined by~\eqref{eq: definition of the task}, at iteration $i$.  Since we are interested in a distributed setting, agents during social learning are only allowed to collect estimators from their local neighborhood $\cN_k$--see Fig.~\ref{fig: multitask diagram}.  

{In the sequel, we show how the formulation~\eqref{eq: multitask optimization problem} and the social learning step~\eqref{eq: social learning step} specialize for regularized  (Sec.~\ref{sec: Regularized multitask estimation}), subspace constrained  (Sec.~\ref{sec: Multitask estimation under subspace constraints}), and clustered (Sec.~\ref{sec: clustered multitask estimation}) multitask estimation.}


\section{Regularized multitask estimation}
\label{sec: Regularized multitask estimation}
{In this section, we focus on the regularization term \( \cR(\cw)\) in~\eqref{eq: multitask optimization problem} and its implications for the learning dynamics. In multitask learning (MTL), regularization is a widely used technique to promote task relationships. 
 In most network applications, the underlying graph structure contains information about the relatedness among neighboring tasks. As such, when considering graph-based MTL applications, incorporating the graph structure into the regularization term is a reasonable and natural step. The smoothness model (under which the tasks are similar at neighboring vertices with the strength of similarity specified by the weight between them) will play a central role in our discussion. This smoothness property is often observed in real world applications (see, e.g., Example~\ref{exple: weather forecasting}) and is rich enough to convey the main ideas behind MTL, as we will see in the sequel.  We will examine two main questions: 1)  How to incorporate  graph-based priors into the regularizer? and 2) How does the resulting MTL algorithm behave? 
}
\vspace{-2mm}
\subsection{Multitask estimation under smoothness}
\label{sec: Multitask estimation under smoothness}
We assume that a symmetric, weighted adjacency matrix $C$ is associated with the connected graph illustrated in Fig.~\ref{fig: multitask network models} (right). If there is an edge connecting agents $k$ and $\ell$, then $[C]_{k\ell}=c_{k\ell}>0$ reflects the strength of the relation between $k$ and $\ell$; otherwise, $[C]_{k\ell}=0$. These weights are usually dictated by the physics of the problem at hand--see, e.g.,~\cite{akyildiz2002wireless},\cite[Ch.~4]{grady2010discrete} for graph construction methods. We introduce the graph Laplacian $L$, which is a differential operator defined as $L=\diag\{C\mathds{1}_N\}-C$. Assuming that the tasks have the same length, i.e., $M_k=M$ $\forall k$, the smoothness of $\cw$ {over the graph is} measured in terms of a quadratic form of the Laplacian~\cite{zhou2004regularization}:\vspace{-1mm}
\begin{equation}
\vspace{-1mm}
\label{eq: smoothness measure}
S(\cw)=\cw^\top\cL\cw=\frac{1}{2}\sum_{k=1}^N\sum_{\ell\in\cN_k}c_{k\ell}\|w_k-w_\ell\|^2,
\end{equation}
where $\cL=L\otimes I_M$ is an extended form of the graph Laplacian \cblue{(defined in terms of the Kronecker product operator $\otimes$)}. The smaller $S(\cw)$ is, the smoother the signal $\cw$ on the graph is. Given that the weights are nonnegative, $S(\cw)$ shows that $\cw$ is smooth if nodes with a large $c_{k\ell}$ on the edge connecting them have similar weight values $\{w_k,w_\ell\}$. Therefore, in order to enforce the prior belief that the target signal $\cw^o=\col\{w^o_1,\ldots ,w^o_N\}$ is smooth with respect to the underlying weighted graph, one may choose in~\eqref{eq: multitask optimization problem}:
\vspace{-2mm}
\begin{empheq}[box={\mymath[colback=blue!5,drop small lifted shadow]}]{equation}
\cR(\cw)=S(\cw),\quad\text{and}\quad\Omega=\mathbb{R}^{MN}
\end{empheq}
Under this choice, {the stochastic gradient algorithm for solving~\eqref{eq: multitask optimization problem} takes the following form:
 \vspace{-1mm}
\begin{equation}
\label{eq: centralized implementation of smoothness}
\bcw_i=\bpsi_i-\mu\eta\cL\bcw_{i-1}
\end{equation}
where $\bcw_i$ is the estimate of $\cw^\star$ at instant $i$, and $\bpsi_i=\col\{\bpsi_{k,i}\}_{k=1}^N$ is the vector collecting the intermediate estimates $\bpsi_{k,i}$ in~\eqref{eq: self-learning step} from across all agents. Since we expect $\bpsi_{i}$ to be an improved estimate compared to $\bcw_{i-1}$, we propose to replace $\bcw_{i-1}$ in~\eqref{eq: centralized implementation of smoothness} by $\bpsi_i$. 
By doing so, we obtain algorithm~\eqref{eq: multitask algorithm}  with the social learning step given by:
\vspace{-2mm}
\begin{empheq}[box={\mymath[colback=beaublue,drop small lifted shadow]}]{equation}
\label{eq: social learning step in case of smoothness}
\bw_{k,i}=\bpsi_{k,i}-\mu\eta\sum_{\ell\in\cN_k}c_{k\ell}(\bpsi_{k,i}-\bpsi_{\ell,i})
\end{empheq}
The substitution of $\bcw_{i-1}$ by $\bpsi_i$ is reminiscent of incremental-type arguments in gradient descent algorithms~\cite{bertsekas1997new}. Analyses in the context of adaptation over networks show that substitutions of this type lead to enhanced network stability since they allow to preserve the stability of the agents after cooperation (see, e.g., \cite[p.~160]{sayed2013diffusion} for details). Regarding algorithm~\eqref{eq: social learning step in case of smoothness}, it follows that, when the spectral radius of the combination matrix $I-\mu\eta\cL$ is equal to one, sufficiently small step-sizes ensuring the individual agents stability will also ensure the network stability\footnote{{\cblue{In this article, a network is said to be stable if  the mean-square-error $\frac{1}{N}\|\cw^\star-\bcw_i\|^2$ converges asymptotically to  a bounded region of the order of the step-size.
}}}~\cite{nassif2018diffusion}.}

\cblue{Proximal based approaches are also proposed in~\cite{wang2018distributed} to solve multitask problems under smoothness. However, these approaches require the evaluation of the proximal operator defined by~\eqref{eq: proximal operator} of the risk $Q_k(\cdot)$ at each~iteration~$i$, which can be computationally expensive.}

 \vspace{-2mm}
\subsection{Bias-variance tradeoff}
\cblue{We next consider the interesting question whether multitask learning is beneficial compared to noncooperation. The answer to this inquiry requires i) studying the performance of algorithm~\eqref{eq: multitask algorithm} relative to the actual agents objectives $\{w^o_k\}$  and then ii) examining when the multitask implementation~\eqref{eq: multitask algorithm} can lead to enhanced performance in comparison to the non-cooperative solution~\eqref{eq: non-cooperative strategy}.} 

\cblue{Algorithm~\eqref{eq: multitask algorithm} was studied in {detail} in~\cite{nassif2018diffusion}. It {was} shown that the network MSD defined by~\eqref{eq: network MSD} is mainly influenced by the sum of two factors, as explained further below.} 
 The first {factor} is the steady-state variance of algorithm~\eqref{eq: multitask algorithm} with respect to the regularized solution $\cw^\star$ in~\eqref{eq: multitask optimization problem}, namely, $\lim_{i\rightarrow\infty}\frac{1}{N}\expec\|\cw^\star-\bcw_i\|^2$. 
The second one is \cblue{the bias or} the average distance between the regularized solution $\cw^\star$ and the unregularized one $\cw^o$, namely, $\frac{1}{N}\|\cw^o-\cw^\star\|^2$. {By increasing {the regularization strength} $\eta$, the variance term is more likely to decrease while the bias term is more likely to increase. 
Understanding this bias-variance tradeoff is critical for understanding the behavior of regularized multitask algorithms.   } 

{We therefore describe in the following the \emph{bias-variance} behavior of algorithm~\eqref{eq: social learning step in case of smoothness} by considering the expressions 
 derived in~\cite{nassif2018diffusion}
 .} These expressions are useful for illustrating the concept of multitask learning. 
 As we will see, instead of involving the vertex domain given by the entries $\{c_{k\ell}\}$ of the adjacency matrix, these expressions involve the graph spectral information defined by the eigendecomposition of the Laplacian $L$.  
 Because the Laplacian is a real symmetric matrix, it possesses a complete set of orthonormal eigenvectors. We denote them by $\{v_1,\ldots,v_N\}$. For convenience, we order the set of real, non-negative eigenvalues of $L$ as $0=\lambda_1<\lambda_2\leq\ldots\leq\lambda_N$, where, since the network is connected, there is only one zero eigenvalue with corresponding eigenvector $v_1=\frac{1}{\sqrt{N}}\mathds{1}_N$~\cite{chung1997spectral}. Therefore, the Laplacian can be decomposed as $L=V\Lambda V^\top$ where $\Lambda=\diag\{\lambda_1,\ldots,\lambda_N\}$ and $V=[v_1,\ldots,v_N]$. 
%
 \vspace{-1mm}
\begin{remark}
\label{remark: performance of smoothness}
Consider \cblue{an MSE} network running the multitask algorithm~\eqref{eq: multitask algorithm} with the second step given by~\eqref{eq: social learning step in case of smoothness}. Assume further that $R_{u,k}=R_u$  $\forall k$ and that $\rho(I-\mu\eta L)\leq 1$. Under these assumptions and for sufficiently small step-sizes and  smooth signal $\cw^o$, it is shown~that~\cite{nassif2018diffusion}
:
\begin{equation}
\vspace{-1.6mm}
\label{eq: steady state variance smoothness}
\lim_{i\rightarrow\infty}\frac{1}{N}\expec\|\cw^\star-\bcw_i\|^2\approx\sum_{m=1}^N\varphi(\lambda_m)
\end{equation}
where \vspace{-1.6mm}
\begin{align}
\label{eq: varphi}
\varphi(\lambda_m)
&=\frac{\mu}{2N}\left(\sum_{k=1}^N[v_m]_k^2\sigma^2_{v,k}\right)\left(\sum_{q=1}^M\frac{\lambda_{u,q}}{\lambda_{u,q}+\eta\lambda_m}\right)
\end{align}
with $\lambda_{u,q}$ the $q$-th eigenvalue of $R_u$ {and $[v_m]_k$ the $k$-th entry of the eigenvector $v_m$}. For the bias term, it can be shown that~\cite{nassif2018diffusion}:
 \vspace{-1.5mm}
\begin{equation}
\vspace{-1.5mm}
\label{eq: bias expression}
\|\cw^o-\cw^\star\|^2=\displaystyle{\sum_{m=2}^{N}}\zeta(\lambda_m),
\end{equation}
where  \vspace{-1.5mm}
\begin{equation}
 \vspace{-1.5mm}
\label{eq: vartheta}
\zeta(\lambda_m)=\left\|\eta\lambda_m\left(R_u+\eta\lambda_mI_M\right)^{-1}\wb_m^o\right\|^2,
\end{equation}
with $\wb_m^o=(v_m^\top\otimes I_M)\cw^o$  the $m$-th subvector  of $\cwb^o=(V^\top\otimes I_M)\cw^o$ corresponding to the eigenvalue $\lambda_m$.\qed
\end{remark}
\noindent For the steady-state variance~\eqref{eq: steady state variance smoothness}, 
observe that it consists of the summation of $N$ terms $\varphi(\lambda_m)$, each one corresponding to an eigenvalue $\lambda_m$ of the Laplacian. The first one $\varphi(\lambda_1=0)$ is independent of the regularization strength $\eta$. 
 The remaining terms $\varphi(\lambda_m\neq 0)$   decrease when $\eta$ increases. Therefore, when  $\eta$ increases, the variance decreases.  From expression~\eqref{eq: bias expression}, we observe that the bias tends to increase by increasing the regularization strength $\eta$. 
 However, an interesting fact arises for smooth  $\cw^o$. To see this, we rewrite the regularizer in~\eqref{eq: smoothness measure} as:
  \vspace{-0.5mm}
 \begin{equation}
 \vspace{-0.5mm}
\label{eq: smoothness measure 2}
S(\cw^o)=(\cwb^o)^\top(\Lambda\otimes I_M)\cwb^o=\sum_{m=2}^N\lambda_m\|\wb_m^o\|^2,
\end{equation}
where we used the fact that $\lambda_1=0$. Intuitively, given that $\lambda_m>0$ for $m=2,\ldots,N$, the above expression shows that $\cw^o$ is considered to be smooth \cblue{over the graph} if $\|\wb^o_m\|^2$ corresponding to large eigenvalue $\lambda_m$ is very small. That is, a smooth $\cw^o$ is mainly contained in $[0,\lambda_c]$, i.e., $\|\wb_m^o\|^2\approx 0$ if $\lambda_m>\lambda_c$, and the smoother $\cw^o$ is, the smaller $\lambda_c$ will be. In this case, the effective sum in~\eqref{eq: bias expression} is over the first $c\ll N$ terms {(corresponding to small eigenvalues $\lambda_m$)} instead of $N$ terms. {We thus conclude that as long as $\cw^o$ is sufficiently smooth, moderate regularization strengths $\eta$ in the range $(0,\infty)$ exist such that the decrease in variance at these values of $\eta$ will dominate the increase in bias. In other words, the} MSD at these values of $\eta$ will be less {than the} MSD at $\eta=0${,} which corresponds to the noncooperative mode of operation. 

{Observe from~\eqref{eq: social learning step in case of smoothness} that the social learning step following from the Laplacian regularization term~\eqref{eq: smoothness measure} involves a single communication step at every stochastic gradient update. When multiple steps are allowed, it is reasonable to expect that performance can be improved. In the following, we show how such solution can be designed.}
\vspace{-2mm}
\subsection{Graph spectral regularization}
\label{sec: graph spectral regularization}
{The main observation behind the introduction of this regularizer is that a smooth $\cw^o$ over a graph exhibits a special structure in the  graph spectral domain (it is mainly contained in $[0,\lambda_c]$, i.e., $\|\wb_m^o\|^2\approx 0$ if $\lambda_m>\lambda_c$)~\cite{smola2003kernels}.  
Graph spectral regularization is} used to leverage more thoroughly the spectral information and improve the multitask network performance. In this case, the network will aim at solving problem~\eqref{eq: multitask optimization problem} with $\Omega=\mathbb{R}^{MN}$ and $\cR(\cdot)$ properly selected in order to promote  the prior information available on the structure of $\cw^o$ in the graph spectral domain.~{The following class of regularization functionals on graphs can be used for this purpose~\cite{smola2003kernels,nassif2019regularization}:}
\vspace{-2mm}
\begin{empheq}[box={\mymath[colback=blue!5,drop small lifted shadow]}]{equation}
\label{eq: graph spectral regularization}
\cR(\cw)=\cw^\top r(\cL)\cw=\cw^\top (r(L)\otimes I_M)\cw
\end{empheq}
where $r(\cdot)$ is some well-defined non-negative function on the spectrum $\sigma(L)=\{\lambda_1,\ldots,\lambda_N\}$ of $L$ and $r(L)$ is the corresponding \cblue{\textit{matrix function} defined as~\cite[p.~3]{higham2008functions}:}
\begin{equation}
\label{eq: graph spectral regularization 1}
r(L)=Vr(\Lambda)V^\top=\sum_{m=1}^Nr(\lambda_m)v_mv_m^\top.
\end{equation}
Construction~\eqref{eq: graph spectral regularization} uses the Laplacian as {a means} to design regularization operators. Requiring {a positive} semi-definite regularizer $r(L)$ imposes the constraint $r(\lambda)\geq 0$ for all $\lambda\in\sigma(L)$. Replacing~\eqref{eq: graph spectral regularization 1} {into}~\eqref{eq: graph spectral regularization}, we {obtain -- compare with the regularizer in~\eqref{eq: smoothness measure 2} to see how an extra degree of freedom is introduced in the multitask network design}:
\begin{equation}
\label{eq: graph spectral regularization 2}
\cR(\cw)=\cwb^\top (r(\Lambda)\otimes I_M)\cwb=\sum_{m=1}^Nr(\lambda_m)\|\wb_m\|^2,
\end{equation}
where $\cwb=(V^\top\otimes I_M)\cw$ and $\wb_m=(v_m^\top\otimes I_M)\cw$. The regularization in~\eqref{eq: graph spectral regularization 2} promotes a particular structure in the  graph spectral domain. It strongly penalizes $\|\wb_m\|^2$ for which the corresponding $r(\lambda_m)$ is large. Thus, one prefers $r(\lambda_m)$ to be large for those $\|\wb_m\|^2$ that are small and vice versa. From the discussion following~\eqref{eq: smoothness measure 2}, it is clear that, under smoothness, the function $r(\lambda)$ must be chosen to be monotonically increasing in $\lambda$. {One typical choice is $r(\lambda)=\lambda^S$  with $S\geq 1$. Example~\ref{example: Graph spectral filtering} further ahead illustrates for instance the benefit of using $\lambda^3$ instead of $\lambda$.}

Assuming the regularizer $r(L)$ in~\eqref{eq: graph spectral regularization} can be written as an $S$-th degree polynomial of the Laplacian $L$, i.e., $r(L)=\sum_{s=0}^S\beta_s L^s$ for some constants $\{\beta_s\}$ (or, equivalently, $r(\lambda)=\sum_{s=0}^S\beta_s\lambda^s$), {and following similar arguments that led to~\eqref{eq: social learning step in case of smoothness}, one arrives at the following social step~\eqref{eq: social learning step}}~\cite{nassif2019regularization}:
\begin{empheq}[box={\mymath[colback=beaublue,drop small lifted shadow]}]{equation}
\label{eq: social learning step in case of spectral regularization}
\left\lbrace
\begin{split}
\bpsi_{k,i}^s&=\beta_{S-s}\bpsi_{k,i}+\sum_{\ell\in\cN_k}c_{k\ell}(\bpsi_{k,i}^{s-1}-\bpsi_{\ell,i}^{s-1}),\quad s=1,\ldots,S\\
\bw_{k,i}&=\bpsi_{k,i}-\mu\eta\bpsi_{k,i}^S
\end{split}
\right.
\end{empheq}
where $\bpsi_{k,i}^0=\beta_S\bpsi_{k,i}$. It requires $S$ communication steps. The resulting {algorithm~\eqref{eq: social learning step in case of spectral regularization}} is distributed since at each step, each agent is only required to exchange information locally with its neighbors. Since $S$ communication  steps are required, agent $k$ ends up collecting information from its $S$-hop neighborhood.

For more general $r(\lambda)$ that are not necessarily polynomial in $\lambda$, one would like to benefit from the sparsity of the graph captured by $L$. As long as  $r(L)$ can be approximated  by some lower order polynomial in $L$, say $r(L)\approx\sum_{s=0}^S\beta_sL^s$, distributed implementations of the form~\eqref{eq: social learning step in case of spectral regularization} are possible--see~\cite{nassif2019regularization}. Problems of this type have already been considered in graph filters design~\cite{shuman2018distributed,sandryhaila2014discrete}.  
For instance, the work~\cite{shuman2018distributed} proposes to \emph{locally} approximate $r(\cdot)$ by a polynomial $\widetilde{r}(\cdot)$ computed by truncating a shifted Chebyshev series expansion of $r(\cdot)$ on $[0,\lambda_N]$. When the regularizer $r(\cdot)$ is continuous, the Chebyshev approximation $\widetilde{r}(\cdot)$ converges to it rapidly as $S$ increases. When the regularizer presents some discontinuities, polynomial approximation methods are not advised over adaptive networks since accurate approximation would require a large order $S$, and consequently, a large number of communication steps at each iteration. Projection based methods {similar to} the one described in Sec.~\ref{sec: Multitask estimation under subspace constraints} can be useful in this case. For example, if the smooth signal $\cw^o$ is only contained in $[0,\lambda_c]$, 
instead of using a discontinuous regularizer $r(\lambda)$ of the form $r(\lambda_m)=0$ if $m<c$ and $\beta\gg 0$ otherwise, one may design a multitask network that is able to project onto the space spanned by the first $c$ eigenvectors of the graph Laplacian.

Since the optimization problems in Sec.~\ref{sec: Multitask estimation under smoothness} and~\ref{sec: graph spectral regularization} are the same with $\cL$ in~\eqref{eq: smoothness measure} replaced by $r(\cL)$ in~\eqref{eq: graph spectral regularization}, the multitask strategy~\eqref{eq: social learning step in case of spectral regularization} will behave in a similar manner as~\eqref{eq: social learning step in case of smoothness}. Particularly, the bias-variance tradeoff discussion continues to apply, and expressions~\eqref{eq: steady state variance smoothness}--\eqref{eq: vartheta} continue to hold with $\lambda_m$ on the RHS of~\eqref{eq: varphi} and~\eqref{eq: vartheta} replaced by the function $r(\lambda_m)$. {
By replacing $\lambda_m$ on the RHS of~\eqref{eq: vartheta} by $r(\lambda_m)$, one may directly observe the consequence of this regularizer on the bias term~\eqref{eq: bias expression}, which can be made now close to zero (by choosing in the smoothness case, for example, $r(\lambda_m)\approx 0$ if $\lambda_m\in[0,\lambda_c]$ and $\beta_m >0$ otherwise).}
\begin{figure*}
\centering
\includegraphics[scale=0.23]{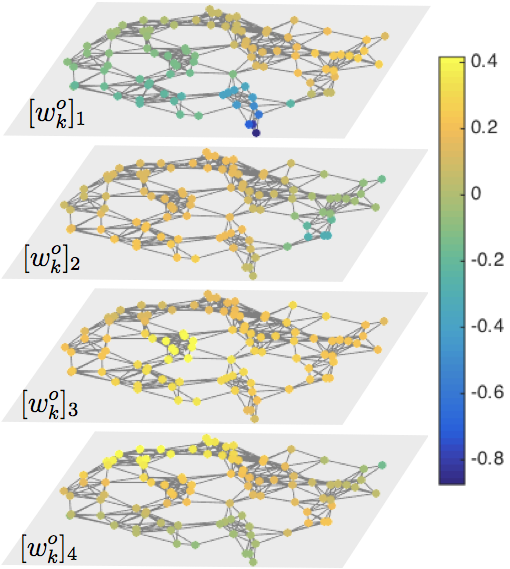}
\includegraphics[scale=0.29]{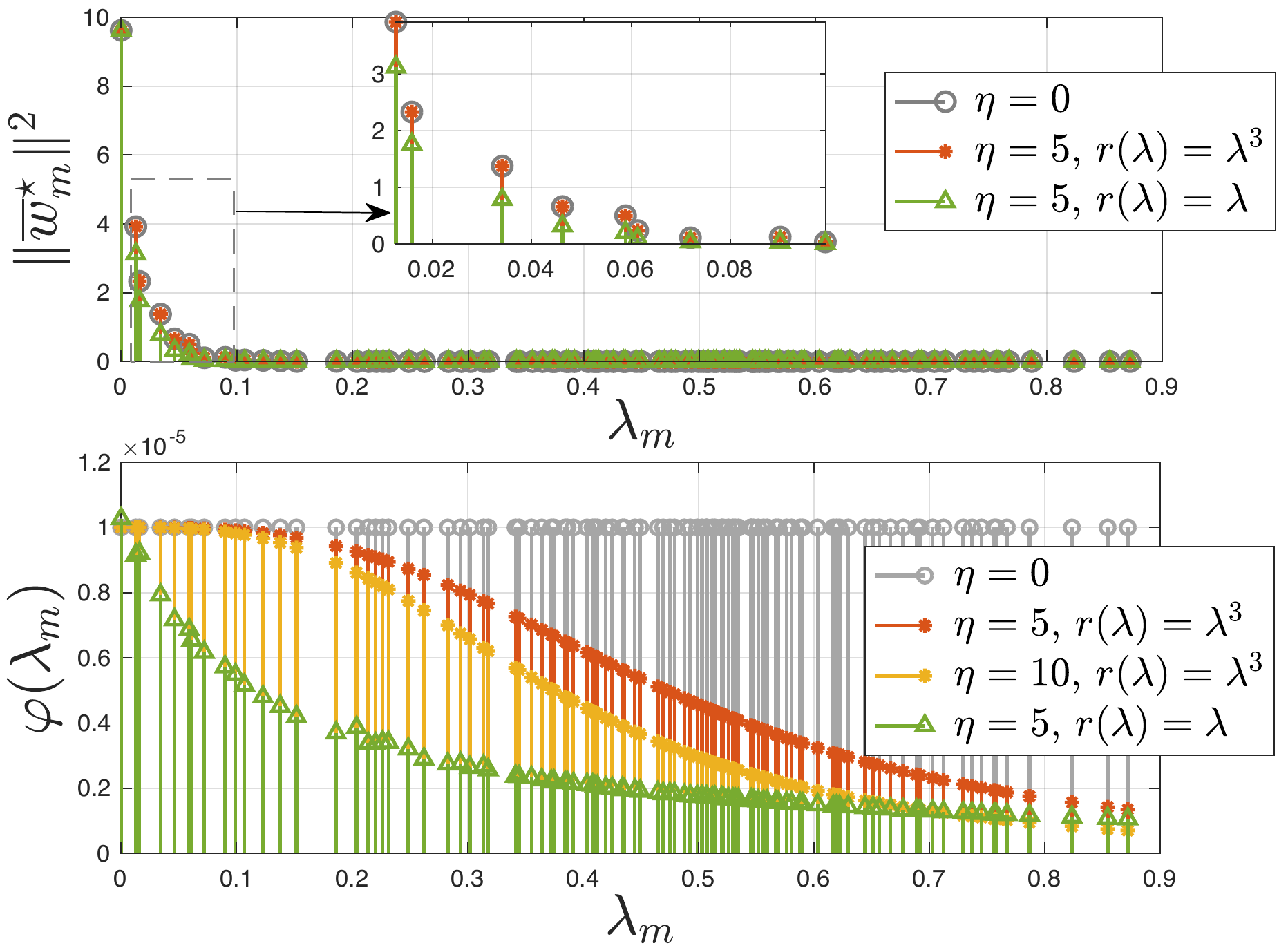}
\includegraphics[scale=0.29]{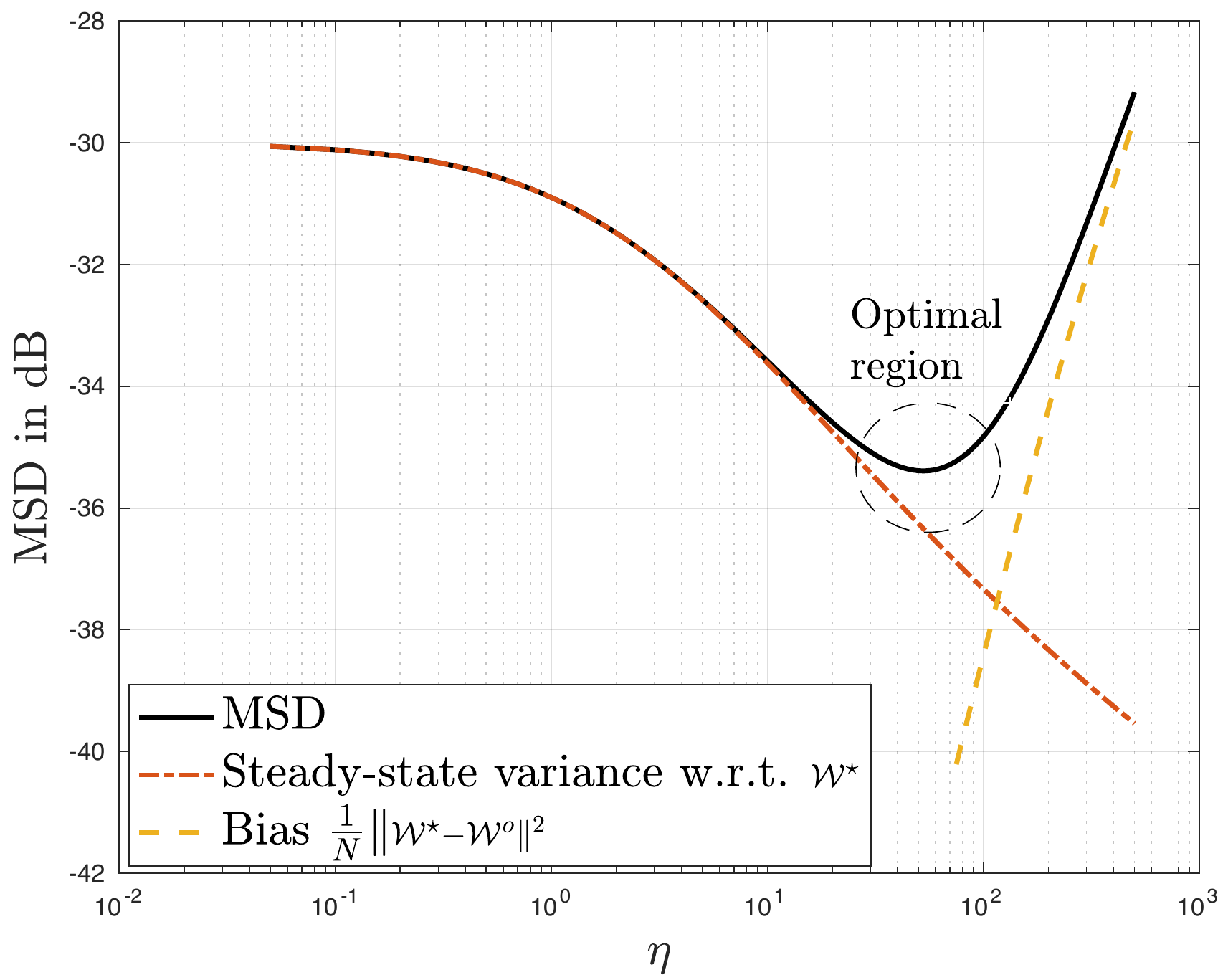}
\vspace{-2mm}
\caption{Illustrative example for spectral regularization. \emph{(Left)} Estimation under smoothness. \emph{(Middle)} Behavior of algorithm~\eqref{eq: multitask algorithm} in the graph spectral domain with $\wb^\star_m=(v_m^\top\otimes I_M)\cw^\star$. \emph{(Right)} Bias-variance tradeoff for $r(\lambda)=\lambda^3$. }
\vspace{-3mm}
\label{fig: example of smoothness}
\end{figure*}
\begin{example}{\emph{(Graph spectral filtering).}}
\label{example: Graph spectral filtering}
\emph{Consider the MSE network example in Fig.~\ref{fig: example of smoothness} and assume \emph{uniform} data profile, i.e., $R_{u,k}=R_u$ and $\sigma^2_{v,k}=\sigma^2_v$ $\forall k$.  In the left plot, we illustrate the entries of the tasks $w^o_k$, which are smooth over the underlying graph. In the middle plot, we illustrate the behavior of the previously described algorithms in the graph spectral domain. The top plot represents the behavior of the network output $\cw^\star$ for three different choices of regularizer $r(\lambda)=\{0,\lambda,\lambda^3\}$. The bottom plot represents the behavior of the steady-state variance~\eqref{eq: varphi} with the eigenvalue $\lambda_m$ replaced by the function $r(\lambda_m)$. Observe how the regularizer $r(\lambda)=\lambda^3$ penalizes low-eigenvalues less than $r(\lambda)=\lambda$, and consequently preserves all the signal components $\overline{w}_m$. Observe further the graph low-pass filtering behavior
~\cite{sandryhaila2014discrete,shuman2018distributed}. 
Small eigenvalues $\lambda_m$ correspond to low frequencies, 
 $\cwb=(V^\top\otimes I_M) \cw$ corresponds to the graph Fourier transform, and $\wb_m=(v_m\otimes I_M)\cw$ corresponds to the $m$-th frequency content of $\cw$. It can be shown that the $m$-th frequency content  of the output   can be bounded as $\|\wb^\star_m\|\leq\frac{\lambda_{u,\max}}{\lambda_{u,\max}+\eta r(\lambda_m)}\|\wb^o_m\|$ in terms of the $m$-th frequency content  of the input $\cw^o$ where $\lambda_{u,\max}$ is the maximum eigenvalue of $R_u$ (see~\cite{nassif2018diffusion})
 .  Since $r(\lambda)$ is monotonically increasing in $\lambda$, for fixed $\eta$, as $\lambda$ increases, the ratio decreases. Therefore, the network output $\cw^\star$ can be interpreted as the output of a low-pass graph filter applied to the signal $\cw^o$. A similar behavior arises for the steady-state variance. For fixed $\eta$, as $\lambda_m$ increases, the variance at the $m$-th frequency, i.e.,  $\varphi(\lambda_m)$, decreases, and  for  fixed $\lambda_m$, as $\eta$ increases, $\varphi(\lambda_m)$ decreases. The regularizer $r(\lambda)$ controls the shape of the filter and the strength $\eta$ controls the sharpness. The non-cooperative solution ($\eta=0$) corresponds to an all-pass graph filter. In the right plot, we illustrate the bias-variance tradeoff in the case of $r(\lambda)=\lambda^3$.}\qed
\end{example}
Returning  to the diagram in Fig.~\ref{fig: multitask diagram}, observe that the \emph{self-learning} step corresponds to the inference step where agent $k$ estimates $w^o_k$ from streaming data $\bx_{k,i}$ and that the \emph{social-learning} step corresponds to the graph filtering step where the agents collaborate in order to perform spatial filtering and reduce the effect of the noise on the network MSD defined by~\eqref{eq: network MSD}. These steps are performed simultaneously. Therefore, multitask learning over networks allows to blend real-time adaptation with graph (spatial) filtering. 
\vspace{-2mm}
\subsection{Non-quadratic regularization}
\label{sec: Non-quadratic regularization}
Non-quadratic regularization has been also considered in the literature
~\cite{nassif2016proximal,hallac2015network,koppel2016proximity}. 
This scenario will induce \emph{non-linearities} in the social learning step~\eqref{eq: social learning step}. In this case, multitask algorithms are derived in order to solve problem~\eqref{eq: multitask optimization problem} with
\vspace{-1.5mm}
\begin{empheq}[box={\mymath[colback=blue!5,drop small lifted shadow]}]{equation}
\label{eq: co-regularizer}
\Omega=\mathbb{R}^{MN} \quad\text{and} \quad \cR(\cw)=\sum_{k=1}^N\sum_{\ell\in\cN_k}\rho_{k\ell}h_{k\ell}(w_k,w_\ell)
\end{empheq}
where $h_{k\ell}:\mathbb{R}^M\times \mathbb{R}^M\rightarrow\mathbb{R}$ is a convex cost function associated with the link $(k,\ell)$. In general, this function  is used to enforce some constraints on the pairs of variables across an edge. {Observe that~\eqref{eq: co-regularizer}, by allowing arbitrary distance measures \( h_{k\ell}(\cdot,\cdot) \), is a generalization of the previously employed quadratic regularization. In fact, setting \cblue{\( h_{k \ell}(w_k, w_{\ell}) = \|w_k - w_{\ell}\|^2 \)} recovers~\eqref{eq: smoothness measure}. Examples of other typical choices are the  $\ell_2$-norm regularizer \cblue{$h_{k\ell}(w_k,w_\ell)=\|w_k-w_\ell\|$} and the $\ell_1$-norm regularizer $h_{k\ell}(w_k,w_\ell)=\|w_k-w_\ell\|_1$. Instead of encouraging global smoothness, these sparsity-based regularizers can adapt to heterogeneity in the level of smoothness of the tasks $w^o_k$ across nodes~\cite{wang2016trend}. Such heterogeneity is observed for instance in the problem of predicting housing prices~\cite{hallac2015network}.  In this problem, the objective  at each node (house) in a graph (where neighboring houses are connected by edges) is to learn the weights $w^o_k$ of a regression model (examples of features are number of bedrooms, square footage, etc.) to estimate the price. Due to location-based factors (such as 
distance to highway) that are often unknown a priori and, therefore, cannot  be incorporated as features, similar houses in different, though close (neighbors), locations can have drastically different prices, i.e., drastically different $w^o_k$. The objective in this case is to encourage neighboring houses that share common models to cooperate without being influenced by the misleading information of neighbors sharing different models, i.e., perform automatic clustering. To do so, the authors in~\cite{hallac2015network} propose to solve the \emph{network Lasso} problem, i.e.,  problem~\eqref{eq: multitask optimization problem} with $\ell_2$-norm regularizer in~\eqref{eq: co-regularizer}. The rationale behind this choice is that  $\ell_2$-norm encourages group sparsity, i.e., consensus across an edge $w_k=w_\ell$.  
On the other hand, the $\ell_1$-norm regularizer  is used in~\cite{nassif2016proximal} to promote the prior that the parameter vectors at neighboring nodes have a large number of similar entries and a small number of distinct entries.} 
The weight $\rho_{k\ell}\geq 0$ in~\eqref{eq: co-regularizer} associated with the link $(k,\ell)$ aims at locally adjusting the regularization strength. It is usually dictated by the physics of the problem at hand. 
For primal adaptive techniques and due to the non-differentiability of the regularizers,  proximal gradient methods can be used to solve~\eqref{eq: multitask optimization problem}. Assuming $\rho_{k\ell}=\rho_{\ell k}$ {and $h_{k\ell}(w_k,w_\ell)=\|w_k-w_\ell\|_1$}, one may arrive to a  multitask algorithm of the form~\eqref{eq: multitask algorithm} with the social learning step~\eqref{eq: social learning step} given by (see the derivations~in~\cite{nassif2016proximal}):
\vspace{-1.5mm}
\begin{empheq}[box={\mymath[colback=beaublue,drop small lifted shadow]}]{equation}
\label{eq: proximal learning step}
\bw_{k,i}=\prox_{\eta\mu \widetilde{g}_{k,i}}(\bpsi_{k,i})
\end{empheq}
where $\prox_{\gamma g}(w')$ denotes the proximal operator of the function $ g(w)$: 
\vspace{-2mm}
\begin{equation}
\label{eq: proximal operator}
\prox_{\gamma g}(w')=\argmin_{w\in\mathbb{R}^M}g(w)+\frac{1}{2\gamma}\|w-w'\|^2,
\end{equation}
and where the function $\widetilde{g}_{k,i}:\mathbb{R}^M\rightarrow\mathbb{R}$ is given by 
$\widetilde{g}_{k,i}(w_k)=\sum_{\ell\in\cN_k}\rho_{k\ell}h_{k\ell}(w_k,\bpsi_{\ell,i}).$
Notice that the proximal operator in~\eqref{eq: proximal learning step} needs to be evaluated at each iteration. For the weighted sum of $\ell_1$-regularizer, a closed form expression can be found in~\cite{nassif2016proximal}. 
%
 %
\vspace{-2mm}
\section{Multitask estimation under subspace constraints}
\label{sec: Multitask estimation under subspace constraints}
{Besides regularized-based algorithms, projection-based algorithms have received considerable attention in the \cblue{literature of deterministic~\cite{lorenzo2019distributed,nedic2009distributed,mota2015distributed,bertsekas1997new} and stochastic~\cite{sayed2013diffusion,sayed2014adaptation,nassif2019adaptation2,alghunaim2019distributed,platachaves2015distributed,sahu2018cirfe} optimization}. The objective in this case is to design distributed networks that are able to project onto low-dimensional subspaces while minimizing the individual costs, i.e.,  solve problems of the form~\eqref{eq: multitask optimization problem} \cblue{with~\cite{nassif2019adaptation2,lorenzo2019distributed}}:}
\vspace{-1.5mm}
\begin{empheq}[box={\mymath[colback=blue!5,drop small lifted shadow]}]{equation}
\label{eq: subspace constraint}
\cR(\cw)=0,\quad\text{and}\quad\Omega=\text{Range}(\cU)
\end{empheq} 
where $\text{Range}(\cdot)$ denotes the range space operator and $\cU$ is an $M_t\times P$ full-column rank matrix with $M_t=\sum_{k=1}^NM_k$ and $P\ll M_t$. {The reader will soon realize that consensus-type problems are instances of this formulation. Also, multitask estimation under smoothness can benefit from this formulation: as explained earlier in Sec.~\ref{sec: graph spectral regularization}, when the first $c$ eigenvectors of the Laplacian are available, the designer can project onto $\text{Range}(\cU)$ with $\cU=[v_1,\ldots,v_c]\otimes I_M$ instead of using regularization.}

 Let $\cP_\cu=\cU(\cU^\top\cU)^{-1}\cU^\top$ denote the projection onto the range space of $\cU$. Assuming that the network topology and the signal subspace $\cU$ are such that the following feasibility problem:
\begin{equation}
\label{eq: feasibility problem}
\begin{array}{cl}
\text{find}&\cA\\
\text{such that} &\cA\,\cU=\cU,\quad\cU^\top\cA=\cU^\top,\quad\rho(\cA-\cP_\cu)<1,\\
&[\cA]_{k\ell}=0,\text{ if  }\ell\notin\cN_k \text{ and }\ell\neq k
\end{array}
\end{equation}
admits at least one solution, one may arrive to a multitask strategy of the form~\eqref{eq: multitask algorithm} with the social learning step~\eqref{eq: social learning step} given by~\cite{nassif2019adaptation2}:
\vspace{-1.5mm}
\begin{empheq}[box={\mymath[colback=beaublue,drop small lifted shadow]}]{equation}
\label{eq: social learning step for subspace constraints}
\bw_{k,i}=\sum_{\ell\in\cN_k}A_{k\ell}\bpsi_{\ell,i}
\end{empheq}
where $A_{k\ell}=[\cA]_{k\ell}$ is the $(k,\ell)$-th block (of size $M_{k}\times M_\ell$) of the $N\times N$ block matrix $\cA$. A matrix $\cA$ satisfying the constraints in~\eqref{eq: feasibility problem} is \cblue{semi-convergent~\cite{nassif2019adaptation2,lorenzo2019distributed}}. Particularly, it holds~that:
\begin{equation}
\lim_{i\rightarrow\infty}\cA^i=\cP_\cu.
\end{equation}
The first two constraints in~\eqref{eq: feasibility problem} state that the $P$ columns of $\cU$ are right and left eigenvectors of $\cA$ associated with the eigenvalue $1$. Together with these two constraints, the third constraint in~\eqref{eq: feasibility problem} ensures that $\cA$ has $P$ eigenvalues at one, and that all other eigenvalues are strictly less than one in magnitude. The last constraint in~\eqref{eq: feasibility problem} corresponds to the sparsity constraint which characterizes the network topology and ensures local exchange of information at each instant $i$. 

{Before explaining how some typical choices of $\cU$ lead to well-studied distributed inference problems,   we note that the  distributed algorithm~\eqref{eq: social learning step for subspace constraints} has an attractive property: in the small step-size regime, the iterates generated by~\eqref{eq: social learning step for subspace constraints} achieve the steady-state performance of the following  gradient projection algorithm~\cite{nassif2019adaptation2}:
\vspace{-1mm}
\begin{equation}
\bcw_i=\cP_\cu\left(\bcw_i-\mu\col\left\{\widehat{\nabla_{w_k}J_k}(\bw_{k,i-1})\right\}_{k=1}^N\right)
\vspace{-1mm}
\end{equation}
which is centralized  since, at each  instant $i$,  agent $k$ needs to send its estimate $\bpsi_{k,i}$ in~\eqref{eq: self-learning step} to a fusion center, which performs the projection, and then sends the result $\bw_{k,i}$ back to the agent. }
\begin{remark}
\label{remark: performance of non-cooperative strategy}
Consider \cblue{an MSE} network running algorithm~\eqref{eq: multitask algorithm} with the social step~\eqref{eq: social learning step} given by~\eqref{eq: social learning step for subspace constraints} with $\cA=[A_{k\ell}]$ satisfying the constraints in~\eqref{eq: feasibility problem}. Assume that the network is seeking $\cw^o\in\emph{Range}(\cU)$. Assume further that $\cU=U\otimes I_M$ where $U=[u_1,\ldots,u_{\bar{P}}]$ is semi-orthogonal
, and that $R_{u,k}=R_u$ and $M_k=M$ for all $k$. Under these assumptions, and for sufficiently small step-sizes, the network \emph{MSD} defined by~\eqref{eq: network MSD} is given by~\cite{nassif2019adaptation2}:
\begin{equation}
\label{eq: projection performance}
\emph{MSD}=\frac{\mu M }{2N}\sum_{m=1}^{\bar{P}}\left(\sum_{k=1}^N[u_m]^2_k\sigma^2_{v,k}\right).
\end{equation}
\end{remark}
\noindent  {
Notice that the projection framework will not induce bias in the estimation. This is because $\cw^o\in\text{Range}(\cU)$, and, therefore, the vector $\cw^\star$ in~\eqref{eq: multitask optimization problem} is equal to $\cw^o$, the network objective. 
Moreover,} the benefit of cooperation can be readily seen by assuming  uniform variances $\sigma^2_{v,k}=\sigma^2_v$ for all $k$. In this case, comparing~\eqref{eq: projection performance} with~\eqref{eq: non-cooperative performance} in the non-cooperative case, we  conclude that $\text{MSD}=(\bar{P}/N) \text{MSD}^\text{nc}$ where $\bar{P}/N\ll 1$. Therefore, the cooperative strategy outperforms the non-cooperative one by a factor of $N/\bar{P}$.


\subsection{Single-task estimation} 
In single-task estimation, the agents are seeking a common  minimizer $w^o$--see Fig.~\ref{fig: multitask network models} (left). {This problem is encountered in many applications. Examples include target localization and distributed sensing (see, e.g.,~\cite{sayed2013diffusion}). Single-task estimation} can be recast in the form~\eqref{eq: multitask optimization problem} {where $\cR(\cdot)$ and  $\Omega$ are chosen according to~\eqref{eq: subspace constraint} with $\cU=\frac{1}{\sqrt{N}}(\mathds{1}_N\otimes I_M)$.}
Several algorithms for solving such consensus-type problems have been proposed  in the literature, including incremental~\cite{bertsekas1997new}, consensus~\cite{nedic2009distributed}, and diffusion\cite{sayed2013diffusion,sayed2014adaptation} strategies. 
Due to lack of space
, we will describe only the class of diffusion strategies, which can be written 
 in the form~\eqref{eq: multitask algorithm} with the social step~\eqref{eq: social learning step}~given~by:
 \vspace{-1.5mm}
\begin{empheq}[box={\mymath[colback=beaublue,drop small lifted shadow]}]{equation}
 \label{eq: social learning step for diffusion}
 \bw_{k,i}=\sum_{\ell\in\cN_k}a_{k\ell}\bpsi_{\ell,i}
 \end{empheq}
where $a_{k\ell}$ 
 corresponds to  the $(k,\ell)$-th entry of an {$N\times N$} doubly-stochastic matrix $A$~satisfying:
\begin{equation}
\label{eq: condition on doubly-stochastic A}
a_{k\ell}\geq 0,~\sum_{\ell=1}^Na_{k\ell}=1,~\sum_{k=1}^Na_{k\ell}=1, \text{and }a_{k\ell}=0 \text{ if }\ell\notin\cN_k.  
\end{equation}
 Several rules for selecting \emph{locally} these combination coefficients have been proposed in the literature, such as the Metropolis rule and Laplacian rule; see, e.g.,~\cite{sayed2014adaptation}. Observe that step~\eqref{eq: social learning step for diffusion} can be written in the form  of~\eqref{eq: social learning step for subspace constraints} with $A_{k\ell}=a_{k\ell} I_M$ and $\cA=A\otimes I_M$, and that  the resulting matrix $\cA$ will satisfy the constraints in~\eqref{eq: feasibility problem}  over a strongly connected network.
 \vspace{-2mm}
\subsection{Multitask estimation with overlapping parameter vectors} 
\label{subsec: Multitask estimation with overlapping parameter vectors}
It is assumed that the individual costs $J_k(\cdot)$ depend only on a subset of the components of a global parameter vector \cblue{$w=[w^1,\ldots,w^M]^\top\in\mathbb{R}^{M\times 1}$~\cite{mota2015distributed,alghunaim2019distributed,platachaves2015distributed,sahu2018cirfe}}. {This situation is observed  in Example~\ref{exple: Power system state monitoring} where  the network global parameter vector is  $w=\col\{w^n\}_{n=1}^{14}$ and where the states $w_k$ to be estimated at neighboring areas partially overlap. 
%
It can be verified that this  problem can also be recast in the form~\eqref{eq: subspace constraint} with $\cU$ properly selected.  To solve this  consensus-type  problem, and motivated by the single-task diffusion strategies, the works~\cite{alghunaim2019distributed,platachaves2015distributed} propose the following algorithm.
}
Assume agent $k$ is interested in estimating the entry $w^n$ of $w$ and let $\cN_k^n$ denote the set of neighbors of $k$ that are interested also in estimating $w^n$. In order to reach consensus on  $w^n$, agent $k$ assigns to its entry $w^n$ a set of non-negative coefficients $\{a_{k\ell}^n\}$ satisfying
\begin{equation}
a_{k\ell}^n=0 \text{ if }\ell\notin\cN_k^n, \quad\sum_{\ell\in\cN_k^n}a_{k\ell}^n =1,\quad\sum_{\ell\in\cN_k^n}a_{\ell k}^n=1,
\end{equation}
and performs the following convex combination: 
\vspace{-1.5mm}
\begin{empheq}[box={\mymath[colback=beaublue,drop small lifted shadow]}]{equation}
\label{eq: rule for the combination when overlapping parameter vector}
\bw^n_{k,i}=\sum_{\ell\in\cN_k^n}a_{k\ell}^n\bpsi^n_{\ell,i}
\end{empheq}
where $\bpsi^n_{\ell,i}$ is the entry of the $M_{\ell}\times 1$ intermediate estimate $\bpsi_{\ell,i}$ (obtained from~\eqref{eq: self-learning step}) corresponding to  the variable $w^n$ and $\bw^n_{k,i}$ is the estimate of $w^n$ at node $k$ and instant $i$. It can also be verified that solution~\eqref{eq: rule for the combination when overlapping parameter vector} can be written in the form
~\eqref{eq: social learning step for subspace constraints} with  the block $A_{k\ell}$ properly selected.

\cblue{For MSE networks, a recursive least-squares (RLS) approach is proposed in~\cite{sahu2018cirfe} to solve overlapping multitask estimation. In general, second-order gradient methods enjoy faster convergence rates than first-order methods at the expense of increasing the~computational~complexity.}
\section{Clustered multitask estimation}
\label{sec: clustered multitask estimation}
Now we move into explaining how clustered multitask estimation can be solved. \cblue{Clustered multitask learning was first considered in~\cite{jacob2008clustered} within the machine learning community. Then, it was extended to adaptation and learning over networks in the work~\cite{chen2014multitask}. As we shall see, clustered multitask estimation} merges subspace constraints with regularization. Let $M_k=M$ for all $k$. In clustered multitask networks, agents within a cluster $\cC_q$ are interested in estimating the same vector $w^o_{\cC_q}$ -- see Fig.~\ref{fig: multitask network models} (middle). Without loss of generality, we index agents according to their cluster indexes such that agents from the same cluster will have consecutive indexes. 
Let $N_q$ denote the number of agents in cluster $\cC_q$. Since agents within $\cC_q$ need to reach a consensus on $w^o_{\cC_q}$, clustered multitask estimation problems can be recast in the form~\eqref{eq: multitask optimization problem} with:
\vspace{-1.5mm} 
\begin{empheq}[box={\mymath[colback=blue!5,drop small lifted shadow]}]{equation}
\Omega=\text{Range}(\cU),\quad\cU=\diag\left\{\frac{1}{\sqrt{N_q}}(\mathds{1}_{N_q}\otimes I_M)\right\}_{q=1}^Q
\end{empheq}
Therefore, the cluster consensus step takes the form~\eqref{eq: social learning step for subspace constraints} with $\cA=A\otimes I_M$ and $A=\diag\{A_q\}_{q=1}^Q$ where the $N_q\times N_q$ blocks  $A_q$ are chosen according to the constraints in~\eqref{eq: feasibility problem}; one typical choice is doubly-stochastic blocks. The resulting $N\times N$ matrix $A=[a_{k\ell}]$ will satisfy:
\begin{equation}
\label{eq: condition on doubly-stochastic A cluster}
a_{k\ell}\geq 0,~A\mathds{1}_N=\mathds{1}_N,~\mathds{1}_N^\top A=\mathds{1}_N^\top, \text{and }a_{k\ell}=0 \text{ if }\ell\notin\cN_k\cap\cC(k),
\end{equation}
where $\cN_k\cap\cC(k)$ denotes the neighboring nodes of $k$ that are inside its cluster. The choice of the regularizer $\cR(\cw)$ in~\eqref{eq: multitask optimization problem} depends on the prior information on how the models across the clusters relate to each other. One typical choice is~\cite{chen2014multitask,nassif2016proximal}:
\vspace{-1.5mm}
\begin{empheq}[box={\mymath[colback=blue!5,drop small lifted shadow]}]{equation}
\cR(\cw)=\sum_{k=1}^N\sum_{\ell\in\cN_k\setminus\cC(k)}\rho_{k\ell}h_{k\ell}(w_k,w_\ell)
\end{empheq}
where  $\cN_k\setminus\cC(k)$ denotes neighboring nodes of $k$ that are outside its cluster and $h_{k\ell}(w_k,w_\ell)$ is a  cost associated with the inter-cluster link $(k,\ell)$. This  function is used to enforce some constraints on the pairs of variables across an inter-cluster edge. Examples are $h_{k\ell}(w_k,w_\ell)=\|w_k-w_\ell\|^2$ to enforce graph smoothness~\cite{chen2014multitask} and  $h_{k\ell}(w_k,w_\ell)=\|w_k-w_\ell\|_1$ to enforce sparsity priors~\cite{nassif2016proximal}.

Clustered multitask algorithms have in general the structure~\eqref{eq: multitask algorithm} with  step~\eqref{eq: social learning step} given by: 
\vspace{-2mm}
\begin{subequations}
\label{eq: clustered multitask algorithm}
\begin{empheq}[box={\mymath[colback=beaublue,drop small lifted shadow]}]{align}
\bphi_{k,i}&=\sum_{\ell\in\cN_k\cap\cC(k)}a_{k\ell}\bpsi_{\ell,i}\label{eq: intra-cluster social learning step}\\[-1.25ex]
\bw_{k,i}&=g'_k(\bphi_{k,i},\{\bphi_{\ell,i}\}_{\ell\in\cN_k\setminus\cC(k)})\label{eq: inter-cluster social learning step}
\end{empheq}
\end{subequations}
{In this algorithm, the \emph{self-learning}} step~\eqref{eq: self-learning step} is followed by an \emph{intra-cluster social learning} step~\eqref{eq: intra-cluster social learning step} where node $k$ receives the intermediate estimates $\bpsi_{\ell,i}$ from its intra-cluster neighbors $\cN_k\cap\cC(k)$ and combines them in a convex manner through the coefficients $\{a_{k \ell}\}$ in~\eqref{eq: condition on doubly-stochastic A cluster} to obtain the intermediate value $\bphi_{k,i}$. The second step~\eqref{eq: inter-cluster social learning step} is an \emph{inter-cluster social learning} step where agent $k$ receives the intermediate estimates $\{\bphi_{\ell,i}\}$ from its neighbors that are outside its cluster $\cN_k\setminus \cC(k)$ and combines them properly using the function $g'_k(\cdot)$ to obtain $\bw_{k,i}$. This step helps to incorporate the available prior information on how the models across the clusters are related into the adaptation mechanism. The function $g'_k(\cdot)$ depends on the regularizer $\cR(\cdot)$. For example, for $\ell_1$-norm co-regularizers $h_{k\ell}$ (with $\rho_{k\ell}=\rho_{\ell k}$), one may arrive to an inter-cluster learning step~\eqref{eq: inter-cluster social learning step} given by 
$\bw_{k,i}=\prox_{\eta\mu \widetilde{g}_{k,i}}(\bphi_{k,i})$
with  $\widetilde{g}_{k,i}(w_k)=\sum_{\ell\in\cN_k\setminus\cC(k)}\rho_{k\ell}h_{k\ell}(w_k,\bphi_{\ell,i})$~\cite{nassif2016proximal}.
\section{Conclusion}
In this article, we explained how prior knowledge about tasks relationships can be incorporated into the adaptation mechanism and how different priors yield different multitask  strategies. It then follows that choosing the optimal strategy for a given problem is equivalent to choosing the task relatedness model which best fits the underlying problem. Choosing the most practically viable solution then balances further this model fit against computational and communication constraints.

There are several other aspects and strategies for multitask learning over graphs that were not covered in this article due to space limitations. For instance, we only focused on multitask networks endowed with parameter estimation tasks. However, distributed detection 
was also considered from a multitask perspective (see, e.g.,~\cite{teklehaymanot2015robust}). 
Online network clustering was also considered. The   objective in this case is to design diffusion networks that are able to adapt their combination coefficients in~\eqref{eq: social learning step for diffusion} in order to exclude harmful neighbors sharing distinct~tasks~\cite{zhao2014distributed,chen2015diffusion}. 
Readers can refer to~\cite{plata2017heterogeneous} to have a list of other literature works that are multitask~oriented.

Multitask learning over graphs is worth exploring further, as there are many potential ideas to build on. For instance, the expressions show the sensitivity of the results to the underlying graph structure. It would be useful to infer the entries $c_{k\ell}$ of the adjacency matrix in~\eqref{eq: social learning step in case of smoothness} simultaneously with the self-learning step~\eqref{eq: self-learning step}. This leads to learning the relations between the tasks simultaneously with the tasks. Automatically determining the optimal regularization strength $\eta$ and allowing  edge regularizers $h_{k\ell}(w_k,w_\ell)$ beyond just the   $\ell_1$-norm constitute also clear extensions. Finally, we believe that the number of multitask learning applications in ``distributed, streaming machine learning'' is vast, and hope to witness increased utilization of the algorithms and theoretical results established in the domain of ``learning and adaptation over networks''.

\bibliographystyle{IEEEbib}
{\linespread{1}\bibliography{reference}}

\end{document}